# Feature Squeezing Mitigates and Detects Carlini/Wagner Adversarial Examples

Weilin Xu, David Evans, Yanjun Qi
University of Virginia
evadeML.org

## ABSTRACT

Feature squeezing is a recently-introduced framework for mitigating and detecting adversarial examples. In previous work, we showed that it is effective against several earlier methods for generating adversarial examples. In this short note, we report on recent results showing that simple feature squeezing techniques also make deep learning models significantly more robust against the Carlini/Wagner attacks, which are the best known adversarial methods discovered to date.

## 1 INTRODUCTION

Adversarial examples are inputs crafted by an adversary to fool a machine learning system. Many techniques for generating adversarial examples have been proposed, and nearly all of these perform some kind of algorithmic search starting from a seed sample with the aim of finding a misclassified example within a distance (according to a given distance metric) limited by an adversarial strength parameter [3, 6, 7, 9].

Feature squeezing is a general method for mitigating and detecting adversarial examples against machine learning models [10]. It reduces the search space available to an adversary by coalescing samples that correspond to many different feature vectors in the original space into a single sample. Feature squeezing is an inexpensive pre-processing technique that can be used with any classification model, so it is orthogonal to other model hardening methods. The main idea of feature squeezing is to compare the model's predictions on both the input sample and a squeezed version of that input sample. For normal samples, with a suitable feature squeezer, both predictions should be very similar; for adversarial examples, the predictions will differ substantially.

Previously, Xu et al. [10] found two simple feature squeezing methods (bit depth reduction and median smoothing) to be extremely effective in mitigating and detecting the adversarial examples generated by Fast Gradient Sign Method (FGSM) [3] and Jacobian-based Saliency Map Approach (JSMA) [7]. In this short note, we evaluate a simple feature squeezing defense against three state-of-the-art adversarial techniques proposed recently by Carlini and Wagner [2]. Our experimental results on both the MNIST and CIFAR-10 datasets show that feature squeezing reduces the success rate of targeted attacks from nearly 100% to below 6% (Table 1). In addition, feature squeezing can effectively detect adversarial examples generated by the Carlini/Wagner methods with 98.80% accuracy on MNIST and 87.50% on CIFAR-10.

We next introduce the threat model and the feature squeezer we used. Section 3 evaluates the robustness of deep learning models enhanced with feature squeezing, and Section 4 reports on our detection experiments.

## 2 BACKGROUND

Carlini and Wagner recently introduced three new gradient-based attack algorithms ($L_2$, $L_\infty$ and $L_0$) that are more effective than all previously known methods in terms of the adversarial success rates achieved with minimal perturbation amounts [2]. Their $L_2$ attack uses a logits-based objective function which is different from all existing $L_2$ attacks, and avoids the box constraint by changing variables. Their $L_\infty$ and $L_0$ are based on the $L_2$ attack and tailored to different distance metrics.

### 2.1 Threat Model

We assume an adversary who has full knowledge of a target model but no ability to influence the model. The adversary's goal is to generate adversarial examples that are misclassified (an *untargeted* attack) or that are classified into a particular class (*targeted* attack).

We use the Carlini and Wagner attack algorithms to generate adversarial examples against the target model. In our evaluation, those adversarial examples are squeezed before being fed to the target model. We do not consider an adversary that adapts to the feature squeezing defense. Such adaptation is an area for further research, but appears to be non-trivial, at least when feature squeezing is used in the proposed detection framework.

### 2.2 Feature Squeezer

There exists many possible feature squeezers in the image space, and both color depth reduction and median smoothing were used in previous work [10]. This short paper focuses on median smoothing with a 2×2 window, because our preliminary results indicated that it is almost always the best squeezer in mitigating Carlini/Wagner adversarial examples.

There are several possible approaches to implement median smoothing with a window size of even numbers. We use the SciPy implementation [8] where the center pixel is always located in the lower right of the 2×2 sliding window. When there are two equal median values (two of the pixels have value $x_1$, and the other two values have value $x_2$), the resulting median value is the larger one ($\max(x_1, x_2)$). Pixels on the edge are padded in the reflect mode as needed.

## 3 ROBUSTNESS EVALUATION

We evaluate all three attack methods, $L_2$, $L_\infty$ and $L_0$ proposed by Carlini and Wagner [2], in both untargeted and targeted variations.

Table 1: **Robustness results for feature squeezing using median smoothing with 2×2 window.** (The first 1000 testing images from each dataset were used in evaluation.)

| Dataset | Carlini's Attack | Targeted Attack | | | | Untargeted Attack | |
|---|---|---|---|---|---|---|---|
| | | Adversary Success Rate | | Accuracy on Adversarial Examples | | Accuracy on Adversarial Examples | |
| | | Original | Squeezed | Original | Squeezed | Original | Squeezed |
| MNIST | $L_2$ | 0.999 | 0.022 | 0.001 | 0.879 | 0 | 0.904 |
| | $L_\infty$ | 1 | 0.011 | 0 | 0.936 | 0 | 0.942 |
| | $L_0$ | 1 | 0.057 | 0 | 0.758 | 0 | 0.817 |
| CIFAR-10 | $L_2$ | 1 | 0.033 | 0 | 0.672 | 0 | 0.682 |
| | $L_\infty$ | 1 | 0.037 | 0 | 0.670 | 0 | 0.661 |
| | $L_0$ | 1 | 0.028 | 0 | 0.702 | 0 | 0.706 |

There are are the most effective adversarial methods found to date. Our defense does not depend on the details of these methods. We note, though, that part of their effectiveness, as measured by the goal of changing as little about the input as possible to achieve the adversary goal of changing the class, is what makes the simple feature squeezing defense so effective against them.

## 3.1 Experimental Design

We use two datasets, MNIST [5] (28×28 gray scale images of handwritten digits) and CIFAR-10 [4] (32×32 color photos of objects) for the experiment. These are the same datasets used in Carlini et al.'s evaluation [2].

Two target models respectively for the two datasets were trained with Carlini's code [1]. We adopted all the default parameters for the three attack methods, except that we changed the maximum iterations of $L_2$ attack from 10,000 to 1000 for consistency with other two methods. We only used the first 1000 testing images in each dataset for generating adversarial examples due to the high computational cost of their methods (around 2 minutes per sample for $L_\infty$ and $L_0$). For each of these images, we generated six different adversarial examples for each image to evaluate the robustness, as there were three different attack methods in two modes (untargeted and targeted).

## 3.2 Results

For both targeted and untargeted attacks, we compare the accuracy on adversarial examples between the original model and the one enhanced with feature squeezing. We also compare the adversary success rate for targeted attacks. The target class is selected by $(l + 1) \mod \#class$, where $l$ is the ground truth class. The success rate of a targeted attack is defined as the proportion of adversarial examples being misclassified as the targeted class.

Table 1 summarizes the results. Feature squeezing using 2×2 median smoothing, barely affects the accuracy on legitimate examples (99.5% to 99.4%) for MNIST, and preserves 93.2% of the original accuracy for CIFAR-10 (73.0% versus 78.3%). But, it dramatically improves the robustness of the two models, reducing the adversary's success rate for targeted attacks from nearly 100% to below 6% for all three attack methods on both datasets. Our defense increases the accuracy on adversarial examples from 0% to at least 81% on MINST and over 66% on CIFAR-10 for all the untargeted attacks, which are over 80% of the original accuracy results on legitimate inputs for both models.

## 4 DETECTING ADVERSARIAL EXAMPLES

Feature squeezing is most effectively used to detect adversarial examples. Using it outside of this framework (that is, just using the output of the classifier on the squeezed input) risks presenting adversaries with new exploit opportunities. The detection framework compares the difference of predictions between an original input and the squeezed version. The score of each input $x$ is computed in $L_1$ norm:

$$\text{score} = |g(x) - g(\text{squeeze}(x))|_1$$

Our intuition is that legitimate examples should have low scores since feature squeezing methods should not to change normal inputs in ways that impact classification. On the other hand, an adversarial example might look dramatically different to a target model with feature squeezing, resulting a large score.

## 4.1 Experimental Design

The detection experiment uses the same design and datasets as the robustness experiments. We used the adversarial examples from the previous experiment (Section 3.1), where each of the first 1,000 legitimate examples was associated with six different adversarial examples. In order to get a balanced dataset for evaluating the detection performance, we used the first 6,000 testing images from each dataset as legitimate samples, resulting in 12,000 images in total. We split the images into two parts: one half as the training set and the rest as the validation set. We selected a threshold value that maximized the detection accuracy on the training set, and report the detection performance measured on the validation set.

## 4.2 Results

Table 2 summarizes the results. The ROC-AUC score of 0.9950 indicates that we obtain a nearly perfect detector on the MNIST dataset. By selecting 0.1147 as the threshold, we get a detection accuracy of 98.80% while the true positive rate is 99.30% and the false positive rate is 1.73%. The actual threshold used by a model operator could be different, depending on the desired trade-off between the false negatives and the false positives.

The detection performance on CIFAR-10 is worse than that on MNIST, where the ROC-AUC score on the validation set is 0.8711

**Table 2: Detection results.** The threshold is selected to maximize the detection accuracy on the training set. All the reported metrics are generated from the validation set.

| Dataset | Validated Detection Performance | | | | |
|---|---|---|---|---|---|
| | ROC-AUC | Threshold | Accuracy | TPR | FPR |
| MNIST | 0.9950 | 0.1147 | 0.9880 | 0.9933 | 0.0173 |
| CIFAR-10 | 0.8711 | 0.7423 | 0.8750 | 0.9527 | 0.2027 |

(compared to 0.9950 on MNIST). We believe that is because the target CIFAR-10 model doesn't have state-of-the-art classification accuracy. Considering the poor accuracy of 78.3% on the first 1000 legitimate examples, the model actually outputs unstable predictions for many legitimate images. As a result, the $L_1$ score could be unusually large with feature squeezing, making it more difficult to distinguish from adversarial examples.

## 5 CONCLUSIONS

Adversarial machine learning is an emerging research area, where our current understanding is mostly limited to empirical results. The experimental success of very simple feature squeezing defenses against complicated adversarial methods does not mean that effective adversarial methods do not exist. Instead, it means we need to rethink the formal definition of an adversarial example and develop a stronger theoretical understanding of the behavior of machine learning models and potential adversaries.

## ACKNOWLEDGMENTS

This work was funded by a grant from the National Science Foundation, and gifts from Amazon and Google. We thank Nicolas Carlini and the other authors for making the source code available to the research community.